\documentclass[epj,final]{svjour}
\usepackage{graphicx}
\usepackage{amsmath}
\usepackage{amssymb}

\begin{document}

\title{Conductance distribution in two-dimensional localized systems
with and without magnetic fields}

\author{J. Prior\inst{1,2}, A. M. Somoza\inst{3}  and M. Ortu\~no\inst{3}}
\institute{Clarendon Laboratory, University of Oxford, Parks
Road, Oxford OX1 3PU, United Kingdom \and 
Departamento de F\'{\i}sica Aplicada, Universidad Polit\'ecnica de Cartagena,
Cartagena 30.202, Spain  \and
Departamento de F\'{\i}sica-CIOyN, Universidad de
Murcia, Murcia 30.071, Spain}
\titlerunning{Conductance distribution in two-dimensional localized systems}

\abstract{
We have obtained the universal conductance distribution of 
two-dimensional disordered systems in the strongly localized limit. 
This distribution is directly related to the Tracy-Widom distribution,
which has recently appeared in many different problems. 
We first map a forward scattering paths model into a problem
of directed random polymers previously solved.
We show numerically that the same distribution also applies
to other forward scattering paths models and to the Anderson model. 
We show that most of the electric current follows a preferential percolation-type
path. The particular form of the distribution depends on the type of leads
used to measure the conductance.
The application of a moderate magnetic field changes the average conductance
and the size of fluctuations, but not the distribution when properly scaled. 
Although the presence of magnetic field changes the universality
class, we show that the conductance distribution in the strongly localized limit 
is the same for both classes.
\PACS{71.23.-k \and 72.20.-i}
} \maketitle

\section{Introduction}

The hypothesis of single-parameter-scaling (SPS) \cite{T74,AA79} 
constitutes the main foundation of
our understanding of localization in disordered systems.
The original formulation of the SPS hypothesis focused on the
average conductance, but it was soon realized that the 
full distribution function of the conductance should be considered.
Then, according to the SPS hypothesis, the conductance distribution
function should depend on a single parameter \cite{shapiro}, for example,
the mean conductance $\langle g\rangle$ or the mean of the logarithm,
$\langle \ln g\rangle$.

The validity of the SPS hypothesis has been thoroughly checked
in one--dimensional (1D) systems, where it has been shown that
all the cumulants of $\ln g$ scale linearly with
system size \cite{Ro92}. Thus,
the distribution function of $\ln g$ approaches a Gaussian
form for asymptotically long systems, which in 1D are always strongly 
localized. It is fully
characterized by two parameters, the mean $\langle \ln g\rangle$
and the variance $\sigma ^2$ of $\ln g$.
In the scaling regime, both parameters are related to each other 
through a universal law,
\begin{equation}
\frac{\sigma^2 \xi}{2L} =1
\label{tau}\end{equation}
where $L$ is the system size and $\xi$ the  localization length, defined as
\begin{equation}
\xi= -\lim_{L \to \infty} \frac{2L}{\langle \ln g\rangle}\, .
\end{equation}
Eq.\ (\ref{tau}) was first derived in Ref.\ \cite{At80} within the 
so--called random phase hypothesis, which assumes that there exists a
microscopic length scale over which phases of complex transmission
and reflection coefficients become completely randomized. 
This relation reduces the two parameters of the distribution to
only one and provides, therefore, a justification and
interpretation for SPS in 1D systems.

The situation in higher dimensional systems
is not as clear as in 1D systems. In those dimensions is far more
difficult to do analytical
calculations and numerical simulations have been limited until
recently to small sample sizes.
In the diffusive regime, although the size of universal conductance 
fluctuations depend on the dimension of the system, the distribution
function of the conductance tends to a gaussian for all dimensions. 
Some people
thought that in the localized regime it could happen something
similar. 
In the strong localization regime, 
$\ln g$ was claimed to be normally distributed in dimensions higher 
than one \cite{CM87,KK92,KM93}.
However, it was pointed out that the distribution is not log-normal 
\cite{Ma02,MM04,MM05,PS06}.
We found numerically that the variance
behaves as \cite{PS05,SP06}
\begin{equation}
\sigma^2=A\langle -\ln g\rangle^{\alpha}+B \label{2D}
\end{equation}
with the exponent $\alpha$ equal to $2/3$ in two-dimensional (2D)
systems.  In three-dimensional systems, preliminary results indicated
a possible value of $\alpha=2/5$ \cite{SP06}, but recent 
more extensive simulations
point out to a value of $1/2$, in agreement with directed polymer 
simulations \cite{KM91}.
The constants $A$ and $B$ are model or geometry
dependent. The precise knowledge of the dependence of $\sigma^2$
with $\langle -\ln g\rangle$ made much easier the numerical
verification of the SPS hypothesis, which we checked for
the Anderson model \cite{PS05}. 

In this paper, we concentrate in  the strongly localized regime in
2D systems.
Although experimental measurements of coherent transport at low
temperatures are difficult in this regime,
knowledge of the conductance
distribution is of interest to better understand variable range
hopping conductance, the metal-insulator transition in three
dimensions or the crossover between the diffusive and the
localized regime in 2D systems. 

In the strongly localized regime, the contribution of each Feynman
path to the tunneling amplitude between two sites decays
exponentially with its distance, and we can expect that some
properties of the conductance (in particular the size
dependence) should be dominated by the shortest or forward-scattering 
paths (FSP).
This approach was introduced by
Nguyen, Spivak and Shklovskii (NSS) \cite{NS85} in a model to account for
quantum interference effects in the localized regime.  Medina and
Kardar \cite{MK92} studied in detail the model. They computed
numerically the probability distribution for tunneling and found
that it is approximately log-normal, with its variance increasing
with distance as $r^{2/3}$ for 2D systems. This is in contrast
with the 1D case, where the variance grows linearly with distance,
and with the implicit assumptions of some works on 2D systems.
In our opinion, this approximation did not receive as much attention
as it deserves, probably because in the SPS regime the localization 
length must be much larger than the lattice
constant, which means that contributions from other paths cannot be
negligible.

A major step forward in the field of random systems was done by 
Tracy and Widom (TW) \cite{TW} who
obtained the distribution function of the largest eigenvalue of 
random matrices belonging to the gaussian ensembles, orthogonal,
unitary and symplectic.
It soon became clear that the TW distribution for the unitary ensemble
also appears in the calculation  of the length
of the longest common subsequence in a random permutation \cite{BD99}
and in many other seemingly unrelated  problems. For an introduction see Ref.\
\cite{majundar}.
Particularly important for our problem is the study by Johansson  \cite{Johansson}
of a specific type of directed polymer model. He obtained the distribution
function of the lowest energy state exactly in terms
of the Tracy-Widom (TW) distribution.
Also relevant for us is the one-dimensional polynuclear growth model \cite{png},
which studies the height fluctuations of a growing interface
and is closely related to the Kardar-Parisi-Zhang equation \cite{KPZ}. 

We showed that for the 2D Anderson model in the strongly 
localized limit the conductance distribution  is related to the TW 
distribution \cite{SO07}.
Here we present results for two different models in the FSP approximation, which is important
to link exact results in directed polymers with the more realistic
Anderson model. 
We show numerically that for a model with disorder with both positive and negative values
(which cannot be exactly mapped to the solvable model) the skewness of the conductance distribution tends without free parameters the TW value. 
The role of percolation versus interference is also analyzed.
We finally study the conductance distribution function in the presence of
a magnetic field. We show that systems in the orthogonal gaussian ensemble 
(Anderson model without magnetic field) and in the unitary gaussian ensemble. 
(Anderson model with magnetic field) tend to the same behavior in the strongly localized limit.

In the next section, we describe the two models that we have used in
our calculations. In section III, we  present a mapping of the
conductance in localized systems to the free energy of directed polymers and 
use Johansson's results to get the distribution function. In sections IV and V, 
we show that the distribution function of the logarithm of the conductance 
in the FSP approximation and in the
Anderson model, respectively, are related to the TW distribution.  
In section VI, we  analyze the effects of a magnetic field. 
We finalize with a discussion and conclusion section.

\section{Model}

We have studied numerically the Anderson model and its 
FSP approximation for 2D samples. 
For the Anderson model, we consider square samples of
size $L\times L$ described by the standard Anderson Hamiltonian
\begin{eqnarray}
H = \sum_{i} \epsilon_{i}a_{i}^{\dagger}a_{i}+ t\sum_{i,j}
a_{j}^{\dagger} a_{i}+ {\rm h.c.} \;, \label{hamil}
\end{eqnarray}
where the operator $a_{i}^{\dagger}$ ($a_{i}$) creates (destroys)
an electron at site $i$ of a cubic lattice and $\epsilon_{i}$ is
the energy of this site chosen randomly between $(-W/2, W/2)$ with
uniform probability. The double sum runs over nearest neighbors.
The hopping matrix element $t$ is taken equal to $-1$, which set
the energy scale, and the lattice constant equal to 1, setting the
length scale. All calculations with the Anderson model are done at
an energy equal to 0.01, to avoid the center of the band.

We have calculated the zero temperature conductance $g$ from the
Green functions. The conductance $g$ is proportional to the
transmission coefficient $T$ between two semi--infinite leads
attached at opposite sides of the sample
\begin{equation}
g= \frac{2e^2}{h}T \label{res}\end{equation} 
where the factor of 2 comes from the spin. 
>From now on, we will measure the conductance in
units of $2e^2/h$. The transmission coefficient can be obtained
from the Green function, which can be calculated propagating layer
by layer with the recursive Green function method \cite{M85,V98}.
This drastically reduces the computational effort. Instead of
inverting an $L^2\times L^2$ matrix, we just have to invert $L$
times $L\times L$ matrices. With the iterative method we can
easily solve square samples with lateral dimension up to $L=400$. 
We have considered ranges of disorder
$W$ equal to 13, 15 and 25, which correspond to localization
lengths of 1.12, 2.4 and 3.2, respectively, and lateral dimensions
up to $L=200$ for the calculation of the distribution function, which
requires a huge number of independent runs to get good statistics 
in the tails. 
For this purpose, we average over a number of  realizations 
larger than $6\times 10^5$ for each disorder and size.
The leads serve to obtain the conductivity
from the transmission formula in a way well controlled
theoretically and close to the experimental situation. 

\begin{figure}
\resizebox{0.45\textwidth}{!}{%
  \includegraphics{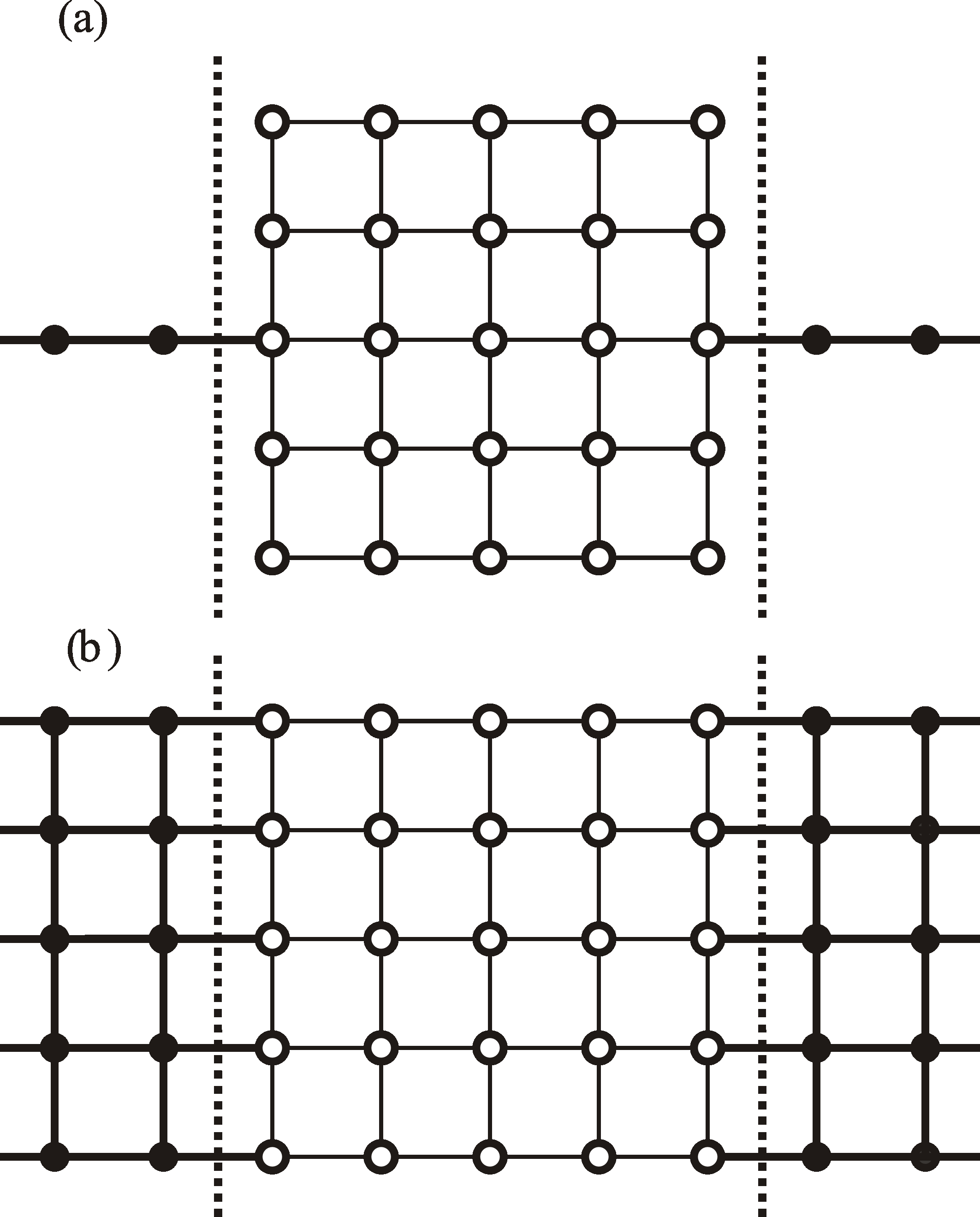}}
\caption{Schematic picture of the sample and the leads considered: (a)
narrow leads and (b) wide leads. The open circles represent sites
in the system and the solid dots sites in the leads. The lines
represent the hopping between sites. The dashed
lines are just a guide to the eye.} \label{f1}
\end{figure}

To consider different possible geometries, we
have used two types of leads: wide leads with the same width as
the lateral dimension of the samples and narrow (one-dimensional)
leads. These are attached to the sample at the centers of opposite
edges, as shown in Fig\ \ref{f1}(a). The scheme of the wide leads
is shown in Fig\ \ref{f1}(b). In both cases the leads are
represented by the same hamiltonian as the system, Eq.\
(\ref{hamil}), but without diagonal disorder. The narrow leads can be
viewed as a simplified model of a point contact, while the wide
leads should roughly correspond to electrodes in contact with the
whole edge of the sample. We use cyclic periodic boundary
conditions in the direction perpendicular to the leads.

The introduction of a uniform magnetic field $B$ perpendicular to
the sample leads to complex hopping matrix elements. In the Landau gauge the vector potential is $\mathbf{A}=(0,-Bx,0)$. Then, the hopping matrix elements
in the $X$ direction are unchanged by the presence of the field,
while the elements in the $Y$ direction have to be multiplied by the factor
$\exp (\pm ixB)$, where the sign depends on whether we are connecting a site with the upper or lower site in the same column \cite{SB84}.

We have also studied the FSP approximation, first considered 
by Nguyen, Spivak and Shklovskii \cite{NS85} and widely studied by 
Medina and Kardar \cite{MK92} in 2D square samples. 
One can write the matrix elements of the Green function between
two sites $a$ and $b$ in terms of the locator expansion
\begin{equation}
\langle a|G|b\rangle
=\sum_{\Gamma}\prod_{i\in\Gamma}\frac{1}{E-\epsilon_i} \;,
\label{mk2}
\end{equation}
where the sum runs over all possible paths connecting the two
sites $a$ and $b$. In general the convergence of this series is
very problematic, but in the strongly localized regime for
distances much larger than the localization length one expect that
the previous sum is dominated by the FSP. Considering only directed paths is well justified in the
strongly localized regime, where the contribution of each
trajectory is exponentially small in its length.
We expect that back-scattering paths renormalize the site energies,
but they should be irrelevant in the renormalization-group sense
in the strongly localized regime. 
Based on this idea, the FSP approximation only considers directed paths.
In this approximation, we will  
consider $E=0$ and two types of diagonal disorder: {\it i}) $\epsilon_i$ can 
only take two values $W$ and $-W$, chosen at random with the same 
probability, which was the model originally considered in Ref.\ \cite{NS85};
{\it ii}) $\epsilon_i=\max [|x|,1]{\rm sign}(x)$, where $x$ is  chosen randomly in the 
interval $(-W/2, W/2)$ with uniform probability. 
The reason to substitute the small disorder energies,
$|\epsilon_i|<1$ by $\pm 1$  is to take partially into account 
the effects of backward paths, which for these sites are important, avoiding at the same time 
problems of convergence in the locator expansion and ensuring that the
transmission through any path is never larger than one. 
We will refer as NSS model 
to the FSP approximation with the first type of disorder.
For the FSP approximation, we concentrate on the transmission amplitude 
between two points in opposite corners of a square sample and we 
assume that the quantum trajectories joining these two points have to follow one of the
(many) shortest possible paths.
For the NSS model it is standard to use the path length $l$, rather 
than the system size $L$. For our geometry, $l=2L$.  
The transmission at zero energy is equal to \cite{MK92}
\begin{eqnarray}
T=\left(\frac{2t}{W}\right)^{2l} J^2(l) \;, \label{mk}
\end{eqnarray}
where the transmission amplitude $J(l)$ is given by the sum over
all the directed paths
\begin{eqnarray}
J(l)=\sum_{\Gamma}^{\rm directed}J_{\Gamma} \;, \label{mk3}
\end{eqnarray}
The contribution of each path, $J_\Gamma$, is the product of the
signs of the disorder along the path $\Gamma$. 
$J$ does not depend on $W$ for the first type 
of disorder and only weakly for the second type, for which
we take $W=10$.

The variance of $\ln g$ is entirely determined by $J^2(l)$ and so 
it depends on $l$. It is convenient to quantify the magnitude of
the fluctuations in terms of the path length
$l$ in the FSP approximation. $\langle \ln g\rangle$ is, of course,
proportional to $l$, but the constant of proportionality depends
on the disorder $W$.
The assumption of directed paths facilitates the computational
problem and makes feasible to handle system sizes much larger than
with the Anderson hamiltonian. The sum over the directed paths can
be obtained propagating layer by layer the weight of all the
trajectories in a very efficient way \cite{MK92}.

\section{Mapping of the conductance in localized systems to polymers}

We have proven that for a specific
FSP model  the distribution 
function of the logarithm of the conductance in 2D systems is a Tracy-Widom
distribution. To do so, we have to map the problem of the
conductance in strongly localized systems to the problem of finding the energy
of a polymer in a random environment. Each FPS path contributing to the conductance
corresponds to a directed polymer.
The calculation of the quantum amplitude between two points, given by Eq. (\ref{mk2}), 
in the FSP approximation is then formally similar to
the calculation of the partition function of directed polymers in
a random potential at a very small temperature $T=1/k\beta$
\begin{equation}
Z =\sum_{\Gamma}\exp\left\{-\beta\sum_{i\in\Gamma}h_i\right\}= \;,
\sum_{\Gamma}\prod_{i\in\Gamma}\exp\left\{-\beta h_i\right\}
\label{pf}
\end{equation}
$h_i$ are random site energies and $\Gamma$
runs over all possible configurations of the directed polymer.
We can easily make Eqns.\ (\ref{mk2}) and (\ref{pf}) equivalent by 
associating $\beta h_i$ with $\ln (E-\epsilon_i)$. In this case, we can map
the distribution of $\ln g$ in our system to the distribution of
the free energy in directed polymers. As in polymer physics
we consider real disorder energies $h_i$, we then choose all the values
$E-\epsilon_i$ to be positive. 

The distribution function of the ground state energy $H$
of a polymer in a disordered environment was obtained exactly by
Johansson \cite{Johansson}. In his model the random
site energies take integer values with probabilities ${\rm
Pr}(h_i=k)=(1-p)p^k$. The ground state energy for
polymers running between the origin and the point $(x,y)$ is given
by \cite{Johansson}
\begin{eqnarray}
 H(x,y)&\rightarrow & \frac{2\sqrt{pxy}+p(x+y)}{1-p}\\
 &+&\frac{(pxy)^{1/6}}{1-p}\left[(1+p)+\sqrt{\frac{p}{xy}}(x+y)\right]^{2/3}\chi_2
\nonumber\label{johan}
\end{eqnarray}
where $\chi_2$ is a random variable with the TW distribution,
corresponding to the distribution of the largest eigenvalue of a
complex hermitian random matrix \cite{TW}. $\chi_2$ verifies 
\begin{equation}
{\rm Pr}(\chi_2 < x)=\int_\infty^x f_2(x') dx'= e^{-g(x)}
\end{equation}
where $g(x)$ is the solution of the equation
\begin{equation}
g''(x)=u^2(x),\label{g1}
\end{equation}
which tends to zero, $g(x)\rightarrow 0$, as $x\rightarrow \infty$. 
$u(x)$ is the
global positive solution of the Painlev\'e II equation
\begin{equation}
u''=2u^3+xu \label{u1}
\end{equation}
which tends to the Airy function, $u(x)\rightarrow {\rm Ai}(x)$,  when $x\rightarrow \infty$.

We can map Johansson's result for the polymer problem
to obtain the distribution function of the conductance for the 
localization problem by considering the FSP
approximation and that the disorder energies are of
the form
\begin{equation}
\epsilon_i=E+e^{\beta k}
\label{mapping} 
\end{equation}
with probability ${\rm Pr}(k)=(1-p)p^k$ for $k=0,1,2,\cdots$, 
in the limit $\beta\rightarrow\infty$. 
This is a very specific model, but we
expect that their results apply in a much more general context,
i.e., different disorder probabilities, as already suggested by 
Johansson \cite{Johansson}.  The same distribution should 
also apply for non zero temperature, as in a 
renormalization-group sense temperature is an irrelevant 
parameter and the behavior is dominated by the 
zero-temperature fixed point.

It is not trivial to extend the previous mapping to realistic problems
with negative site energies, like the NSS and the Anderson model.
It is then natural to study if the distribution function given by
Eq.\ (\ref{johan}) also applies to these models.

\section{Conduction in the forward scattering paths approximation.}

\begin{figure}[htb]
\resizebox{0.50\textwidth}{!}{%
  \includegraphics{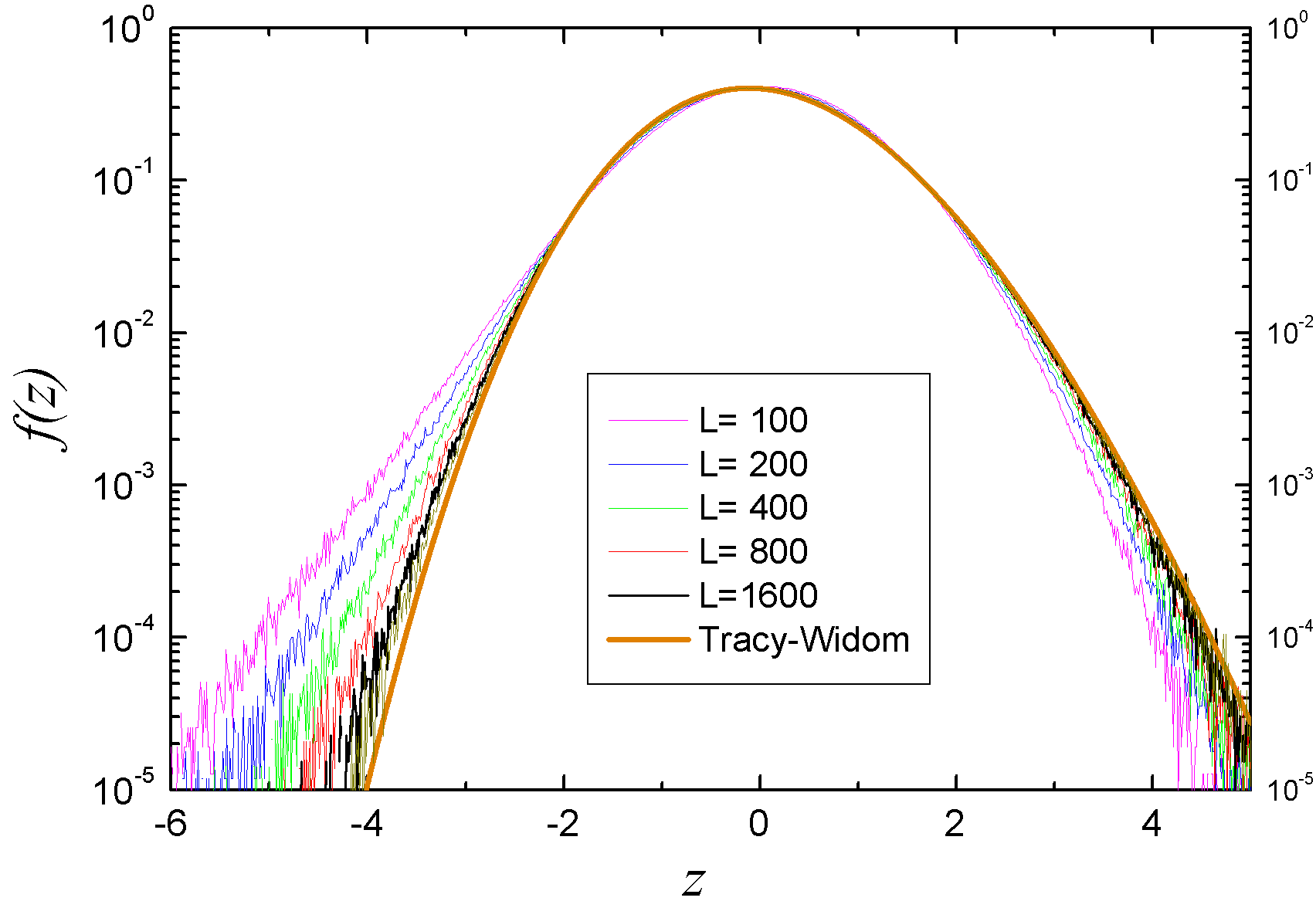}}
\caption{Histograms of $\ln J^2$ as a function of the standardized
variable $z$ for several sizes of the NSS model. 
The thick solid line corresponds to the TW distribution.
As size increases, the data approach the TW distribution.} \label{fig3}
\end{figure}

First of all, we have studied the conductance distribution of the 
NSS model in order to check if it is also given by the TW function. 
For this model it is natural to consider $\ln J^2$, which contains
all the relevant information about fluctuations.
In Fig.\ \ref{fig3} we plot, for several system sizes, histograms 
of $\ln J^2$  as a function of $z=(\ln J^2-\langle \ln J^2\rangle)/\sigma$, 
where $\sigma^2$ is the variance of $\ln J^2$. 
The data are for narrow leads and each line
corresponds to a different system size, specified in the figure.
The thick solid line is the standardized TW distribution.
We see how the results for the NSS model tend uniformly to the TW distribution as size 
increases, although the convergence is relatively slow.

\begin{figure}[htb]
\resizebox{0.50\textwidth}{!}{%
  \includegraphics{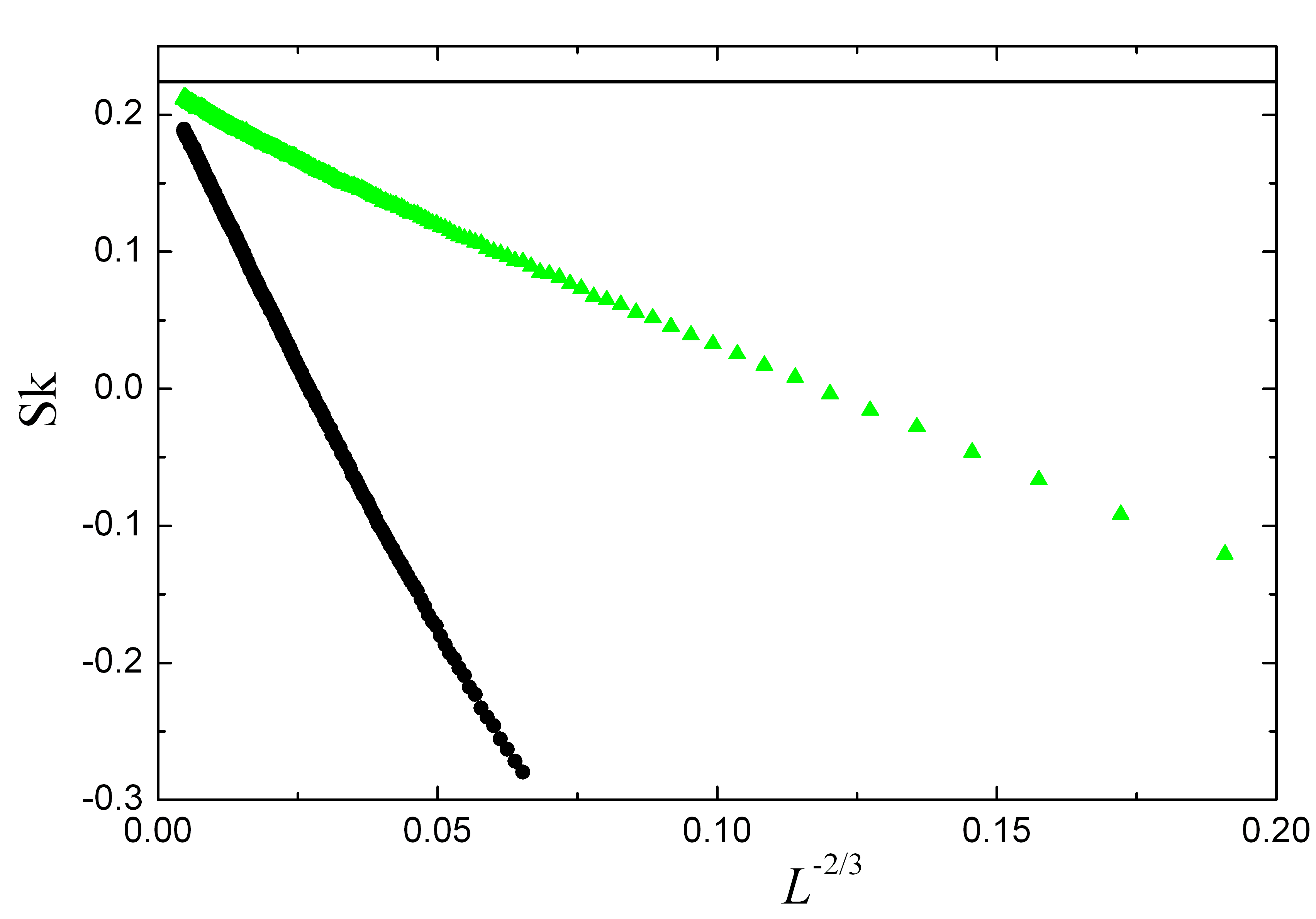}}
\caption{Skewness vs $L^{-2/3}$ for the NSS model (black circles) 
and the FSP approximation with continuous disorder (green triangles). 
The horizontal line corresponds to the TW value.} \label{fig5}
\end{figure}

Although the results in Fig.\ 2 present noticeable finite size effects, 
it is clear that the TW distribution is fully consistent with the limiting distribution. This fact together with the previous mapping permit us to suggest that
\begin{equation}
\ln J^2  = \alpha L + \beta L^{1/3} \chi_2\label{fff}
\end{equation}
in the limit $L\rightarrow \infty$.  
To better guarantee that the TW distribution is the proper limiting function, it is convenient to focus on the adimensional parameters of the distribution, like the skewness (${\rm Sk}=k_3/k_2^{3/2}$) or the kurtosis (${\rm Kur}=k_4/k_2^2$). If the previous equation is correct, these two parameters must tend to the TW values as $L\rightarrow \infty$ independently of the values of $\alpha$ and $\beta$. From Eq.\ (\ref{fff}) we expect that the leading correction for
the skewness, with respect to the limiting value at $L\rightarrow\infty$, is proportional to  $L^{-2/3}$.
In Fig.\ \ref{fig5} we plot the skewness as a function of $L^{-2/3}$ for the NSS model (circles) and for our second FSP model, with continuous disorder (triangles). The horizontal line corresponds to the TW value.
We see how the skewness approaches the correct value and how it does so
in the expected way for the two FSP models considered. 
Finite size effects for the NSS model are larger
than for the other FSP model considered. A similar behavior is also observed 
for the kurtosis. It is clear from these trends that both models tend 
to the same universal distribution.

\begin{figure}[htb]
\resizebox{0.40\textwidth}{!}{%
  \includegraphics{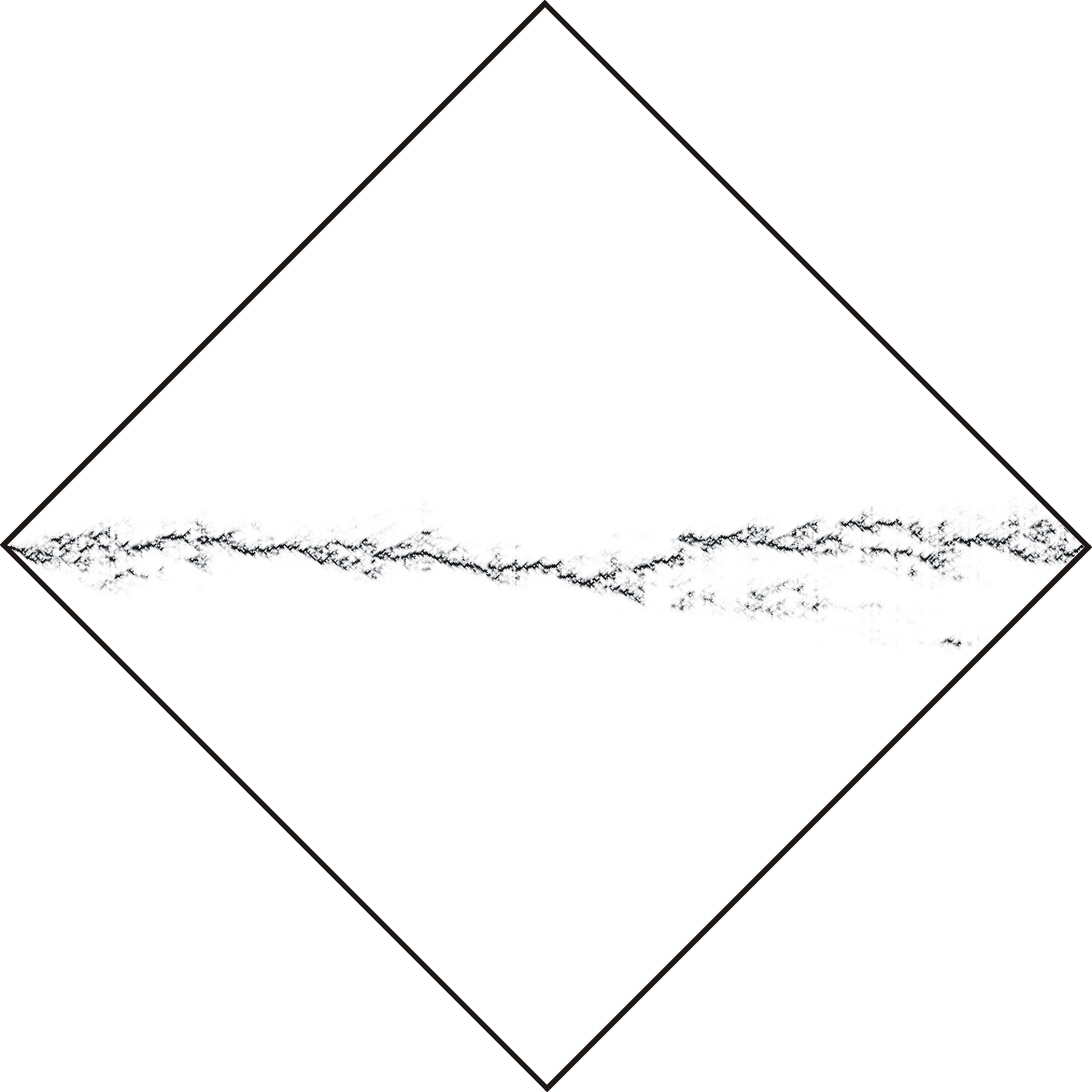}}
\caption{Plot of $J^2$ from the left corner to any other point in a
given sample on a grey scale. On each vertical slice, black (white)
corresponds to the maximum (minimum) conductance.}
\label{foto1}
\end{figure}

We note that Johansson's results imply that the conductivity, 
in our case, is likely to be dominated by the most important FSP. 
On the contrary, the NSS model 
was designed to maximize the interference effects and all trajectories 
have the same amplitude. So, considering only the most conducting path has no sense in this case.
It is necessary then to understand how the NSS may end up reproducing the TW distribution. 
In Fig.\ \ref{foto1} we plot, for a given sample, the value of $J^2$ 
from the left corner to any other point in the sample.  
All points in a vertical slice are at the same hopping distance 
from the left corner. For each vertical slice the maximum value of 
$J^2$ is plotted in black and the minimum in white. 
Intermediate values are plotted on a gre
y scale proportional to $J^2$.
This method gets rid of the exponential decay of $J^2$ with distance.
However, it does not take into account that the current must flow through the
end point.
Despite the fact that all single trajectories have exactly the same 
weight, interference effects produce that most of the current is 
carried through very few well defined paths. The interference effects 
play a role only at short scales, producing a kind of renormalization 
process, which ends up in a dominant ``renormalized'' path. 
Probably, this renormalization also explains why finite size effect are larger in the NSS model than in the other FSP model considered. 

\section{Conductance for the Anderson model}

\begin{figure}[htb]
\resizebox{0.50\textwidth}{!}{%
  \includegraphics{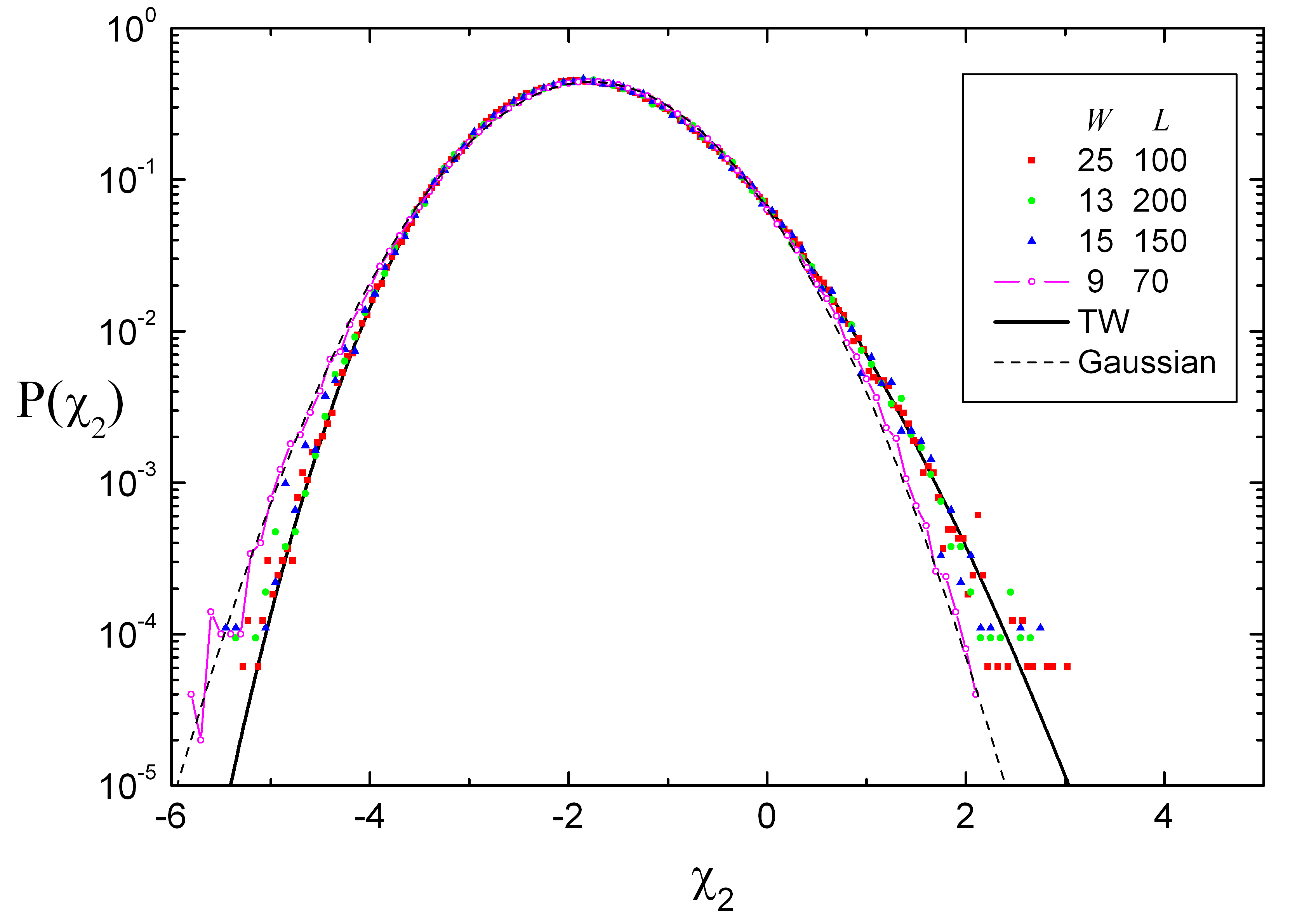}}
\caption{Histograms of $\ln g$ versus the scaled
variable $\chi$ for several sizes and disorders of the Anderson
model with narrow  leads.
The continuous line corresponds to the TW distribution and the dashed line
is a gaussian with the same mean and variance as the TW distribution.}
\label{fig2a}
\end{figure}

Once we have shown that the conductance in the FSP approximation follows
the TW distribution, we study the
Anderson model, which cannot be directly mapped to the polymer problem.
We have calculated numerically the conductance for the 2D Anderson model 
and we have obtained its distribution function.
By similarity with the FSP models we first consider the narrow leads geometry. 
In Fig.\ \ref{fig2a} we plot histograms of $\ln g$ for this model as a function
of $\chi=(\ln g-A)/B$, where $A$ and $B$ are chosen in order to have the same mean and variance
as the theoretical distribution,  the TW distribution $f_2(\chi_2)$. 
The data are for  several sizes and ranges of the disorder, $W=25$ and $L=100$ 
(squares), $W=13$ and $L=200$ (dots), $W=15$ and $L=150$ (triangles) and
$W=9$ and $L=70$ (connected circles).
The solid line corresponds to $f_2(\chi_2)$. 
All the cases in the strongly localized regime, represented by unconnected solid symbols, 
are fitted fairly well by $f_2(\chi_2)$. The agreement extends over more than four 
orders of magnitude. 
The connected empty circles correspond to a system near the crossover 
regime, whose distribution is close
to a log-normal, represented by a dashed line in Fig.\ \ref{fig2a}.
For narrow leads, $f_2(\chi_2)$ fits
the data better than the gaussian function for $L/\xi\gtrsim 13$.

Considering the size dependence of the mean and the variance of
$\ln g$ and the excellent agreement between our data and the TW
distribution, we conclude that in the strongly localized regime
with narrow leads
\begin{equation}
\ln g=-\frac{2L}{\xi}+\beta
\left(\frac{L}{\xi}\right)^{1/3}\chi_2 \label{imp}\end{equation}
where  $\beta$ is a new constant and $\chi_2$ a random variable with
the TW distribution.
In the SPS regime $\beta$ is a  constant, independent of the
disorder, the system size or the Fermi energy. We found, from
present results and previous calculations on the behavior of the
variance, that it is approximately equal $3.4$ for the Anderson
model with narrow leads. Eq.\ (\ref{imp}) must be also valid
outside the SPS regime (when the localization length is very small
or when the Fermi energy lies in the band tails), but with a
non universal value of $\beta$. The data set in Fig.\ \ref{fig2a} for $W=25$
is outside the SPS regime, since it corresponds to a localization 
length of the order of the lattice spacing ($\xi=0.97$), and fits the 
TW distribution very well. The value of $\beta$ is 3.5 in this case.

\begin{table}
\caption{Mean, variance, skewness, and kurtosis for the
distributions  $f_2$ and $f_0$.}
\begin{tabular}{ccc}
&$f_2$&$f_0$\\
\noalign{\hrule}
Mean&-1.77109&0\\
Variance&0.81320&1.15039\\
Skewness&0.2241&0.35941\\
Kurtosis&0.09345&0.28916\\
\end{tabular}
\end{table}

In table I we present the characteristic parameters of the distribution
of $f_2(\chi_2)$.
The fact that it has a mean different from zero,
 $\langle \chi_2\rangle=-1.77109$, has some practical consequences.
According to Eq.\ (\ref{imp}), this implies a contribution to
$\langle \ln g\rangle$ proportional to
 $L^{1/3}$, already observed by us \cite{PS06} in the Anderson 
 model with narrow leads.
 This  is of practical interest for the calculation of the
localization length.  Neglecting this term can
cause errors of the order of 20 \%\ in the estimate of the
localization length.

\begin{figure}[htb]
\resizebox{0.50\textwidth}{!}{%
  \includegraphics{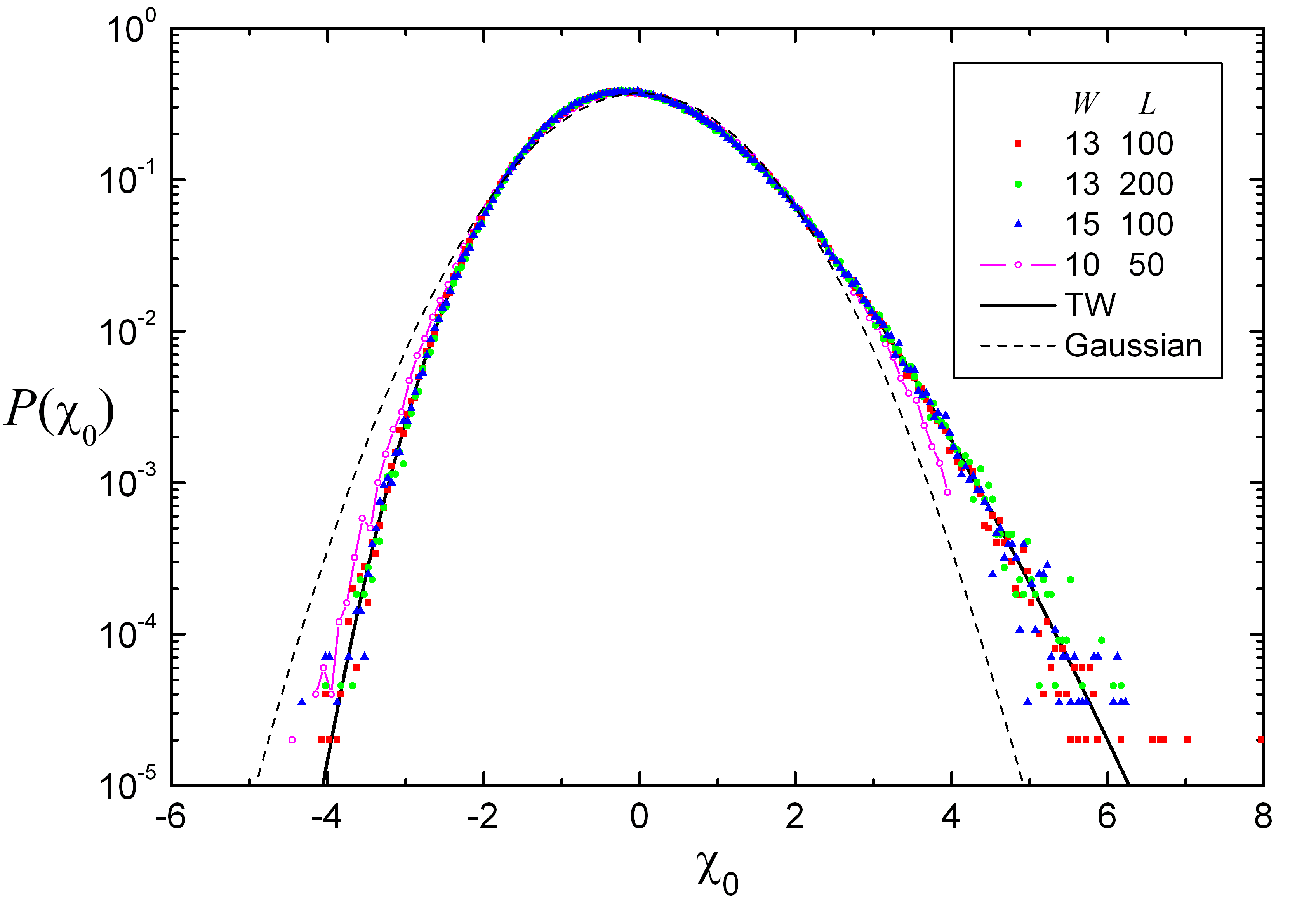}}
\caption{Histograms of $\ln g$ versus the scaled
variable $\chi$ for several sizes and disorders of the Anderson
model with  wide  leads.
The continuous line corresponds to  $f_0$ and the dashed line
is a gaussian with the same mean and variance as $f_0$.}
\label{fig2b}
\end{figure}

We did not find any $L^{1/3}$ contribution for wide leads,
which constitutes a strong indication that the conductance
distribution may depend on the leads, even in the strongly
localized regime. The type of leads may change the universality class.
We can expect a situation similar to polynuclear
growth models \cite{png}, where the height distribution was found
to depend on the initial conditions. Our problem with narrow leads
is directly related to the droplet model in Ref.\ \cite{png}, which
starts from an initial preferential point. With wide leads we have
translational invariance and all the initial (and final) points
are equivalent, a problem similar to stationary growth. In this
case, the height fluctuations are described by the function $f_0$
whose accumulated distribution is \cite{png}
\begin{equation}
\int_\infty^x f_0(x')dx'=\left[1-(x+2 f''+2 g'')g'\right] e^{-(g+2f)}
\end{equation}
where  $f(x)$ is the solution of the equation
\begin{equation}
f'(x)=-u(x),
\end{equation}
which tends to zero when $x \rightarrow
\infty$.  $g(x)$ and $u(x)$ are given by Eqs.\ (\ref{g1}) and
(\ref{u1}), respectively. 
The characteristic parameters of the
distribution $f_0$ are given in table I.
As required by our previous finding that there were no $L^{1/3}$ 
contributions to $\langle\ln g\rangle$, $f_0$ has zero mean. 
In Fig\ \ref{fig2b} we show the histograms of $\ln g$
for the Anderson model with wide leads for several disorders and
sizes, $W=13$ and $L=100$ 
(squares), $W=13$ and $L=200$ (dots), $W=15$ and $L=100$ (triangles) and
$W=10$ and $L=50$ (connected circles). The solid symbols correspond to systems in the
strongly localized regime, while connected empty symbols to a case near the crossover.
We plot in terms of  $\chi=(\ln g-A)/B$, where $A$ and $B$ 
are chosen in order to have the same mean and variance
as the distribution $f_0$.
The full line  corresponds to 
$f_0$ and the dashed line to a gaussian with the same mean and variance. 
The agreement between the numerical data
and the theoretical distribution is again excellent, showing that
$\ln g$ satisfies in this case
\begin{equation}
\ln g=-\frac{2L}{\xi}+\beta \left(\frac{L}{\xi}\right)^{1/3}\chi_0
\label{imp2}\end{equation} where $\beta$ is a new constant and
$\chi_0$ a new random variable with distribution $f_0(x)$. 
In the SPS regime, $\beta$ is a universal constant approximately equal
to 2.2.  
We note that, in the data for $W=10$ and $L=50$, the right tail 
of the distribution is cut at $\chi_0\approx 4$. This is due to the 
existence of a kink in the distribution for a transmission equal to one. 
For wide leads the distribution $f_0(\chi_0)$ fits the data well when
$L/\xi\gtrsim 6$.

Our results support the idea that the Anderson  model and its FSP approximation in 2D
systems in the strongly localized regime will verify
an equation of the form (\ref{imp}) or (\ref{imp2}) for any range
of parameters, type of disorder, geometry or boundary conditions.
The distribution  of the random variable will depend on boundary conditions,
as we have shown for two particular cases, narrow and wide leads. 
It becomes clear that the definition of a ``universality class" requires
to take into account the boundary conditions. 
A detailed study of these was done in the context of a PNG model  \cite{png}.
By analogy with this model, we think that the concept of universality class remains valid
as it seems that there exist only a finite set of universal distributions.
It is difficult to know the complete set of universal distributions
and to which of them will tend a general complex type of lead.

Despite the existence of several universality classes, there are 
some results which are very robust, like the localization length 
and the exponent of $1/3$ in Eqs.\ (\ref{imp}) and
(\ref{imp2}) characterizing the size of fluctuations. This exponent
determines the behavior of the cumulants and
the tails of the distribution \cite{zhang}.
For the two
types of leads studied we have
\begin{equation}
-\ln f_j(x)=\frac{d_jx^{3/2}}{3}\qquad \mbox{ for}\quad
x\rightarrow\infty
\label{cumul}
\end{equation}
 with $d_0=2$ and $d_2=4$. Eq.\ (\ref{cumul})
is valid for the high conductance tail only. The other tail might
be more sensitive to boundary conditions, although both
distributions $f_0$ and $f_2$ behave as \cite{png} 
\begin{equation}
-\ln f_j(x)=\frac{|x|^{3}}{12}\qquad \mbox{ for}\quad
x\rightarrow -\infty.
\end{equation}

As we have mentioned, we consider $L/\xi \gtrsim 6$ as the condition
for the strongly localized regime, where the conductance
distribution is fitted by  Eqs.\ (\ref{imp}) and (\ref{imp2}) much
better than by  a log--normal. For mesoscopic samples close to the
condition  $L/\xi \approx 6$ it should be possible to test
experimentally our predictions, since $g\approx g_0 \exp(-12)$
where $g_0$ is close to one in units of $2 e^2/h$. Our results can
also be verified experimentally through the behaviour of the cumulants of the
distribution. Eqs. (7) and (9) predict universal values for the
skewness, kurtosis, etc, of the distribution, given in table I. 
This limiting values
can be obtained from a measurement of the cumulants in any range
of parameter in the localized region, since Eq.(6) it is fairly
well verified even near the crossover. From the tendency of
the second and third cumulants, for example, it is possible to derive the
asymptotic value for the skewness, which for wide
leads should be $A_3/A_2^{3/2}=0.359$ (see table I).

\section{Distribution function with a magnetic field}

We have studied the distribution function of $\ln g$ in the localized regime of 
the Anderson model at zero temperature when we
apply a magnetic field perpendicular to the sample. 
The field changes the mean, producing negative magnetoresistance, and the variance 
of the distribution but not the distribution itself
when it is rescale in terms of the variable $\chi=(\ln g-A)/C$, 
where $A$ and $C$ are again chosen in order to have the same mean and variance
as the theoretical distributions $f_2$ and $f_0$.
We found that the distribution functions for both narrow and wide leads are the same
as without a magnetic field, Eqs.\ (\ref{imp}) and (\ref{imp2}). 

\begin{figure}[htb]
\resizebox{0.50\textwidth}{!}{%
  \includegraphics{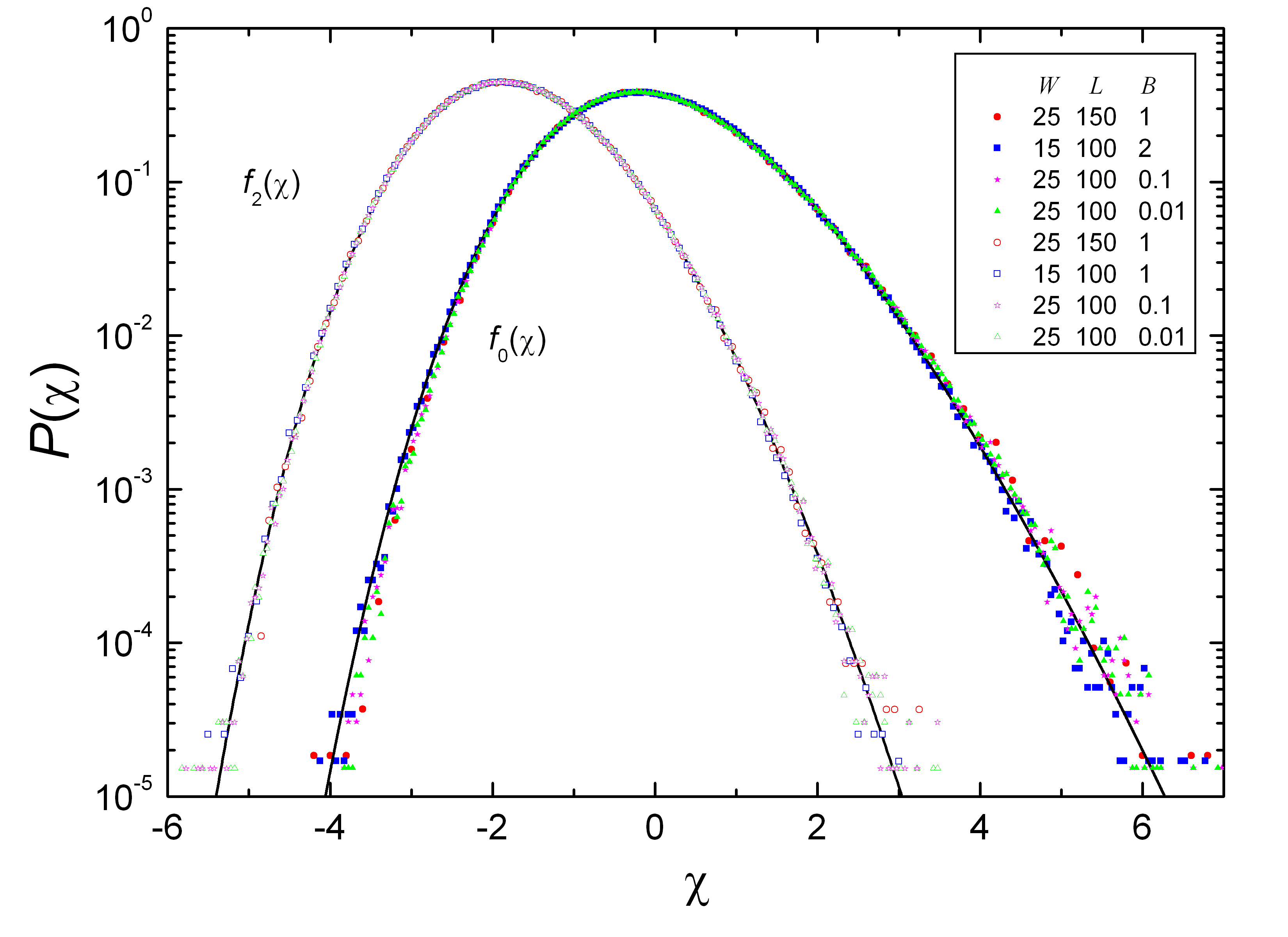}}
\caption{Histogram of $\ln g$ for the Anderson
model with an applied magnetic field for narrow (empty symbols) and 
wide (solid symbols) leads for the values of the disorder, system size and magnetic field given in the figure.
The continuous
lines correspond to $f_2(\chi_2)$ and $f_0(\chi_0)$.} \label{fig6}
\end{figure}

In Fig.\ \ref{fig6} we plot the distribution function of $\ln g$ for 
two values of the disorder $W=15$ and $W=25$, two system sizes and
several values of the magnetic field $B$ (see the legend in the figure). 
We have considered the two types of leads used before: solid symbols correspond 
to wide leads and empty symbols to narrow leads.  The solid lines are our 
theoretical distributions $f_2(\chi_2)$ (narrow leads, left curve) and $f_0(\chi_0)$ 
(wide leads, right curve). We can see that all the points are fitted 
pretty well by the  distributions  $f_2$ and $f_0$,
respectively, as in the absence of a magnetic field. 
We note that we have used periodic boundary conditions
through out the paper, including this section.

The results indicate that the magnetic field does not change the percolative nature
of the conduction in the strongly localized regime.

The value of $\beta$ in Eqs.\ (\ref{imp}) and (\ref{imp2}) changes 
very little with the applied field, less than 
our uncertainty in the measurements. 

We have also checked that the application of a magnetic field in the
NSS model with narrow leads produces negative magnetoresistance, but
does not change the distribution function. To be quantitative, we have
plotted in Fig.\ 8 the skewness as a function of $L^{-2/3}$ for several 
values of the magnetic field, detailed in the legend of the figure. 
The horizontal line corresponds to the 
skewness of the distribution $f_2$ and the lower curve to the situation
in the absence of a magnetic field.
The skewness clearly tends to the TW value, as the system size tends 
to infinity, for all cases considered. We note that in the presence of a field
the  convergence is faster than in its absence. 
It is also interesting to note the dependence of the slopes in Fig.\ 8
with magnetic field. They increase with the field, while the value in the
absence of a field is larger than in the presence of it. 

\begin{figure}[htb]
\resizebox{0.50\textwidth}{!}{%
  \includegraphics{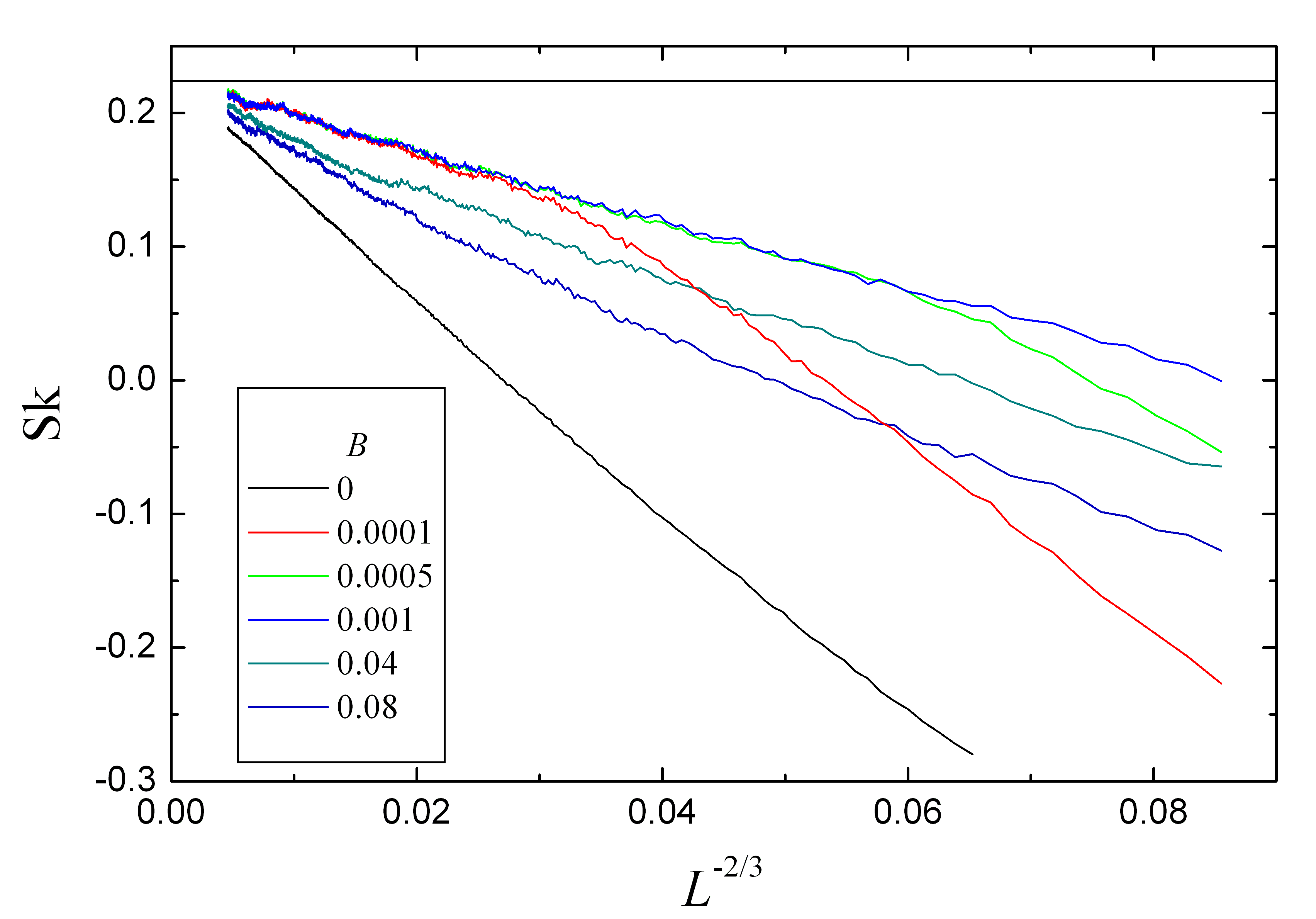}}
\caption{Skewness as a function of $L^{-2/3}$ for several
values of the magnetic field in the NSS model with narrow leads. 
The horizontal line corresponds to the TW value.} \label{fig9}
\end{figure}

\section{Discussion and conclusions}

Present results confirm our previous belief that, in the strongly
localized regime, directed path models are in the same
universality class as the Anderson model \cite{SP06,NS85,MK92}.
While the NSS model pretended to maximize interference effects,
Johansson's model only considers the most important path. The
agreement between both models indicates that it is percolation and
not interference the dominant effect in this regime. We expect
that the main effect of interference between different paths is a
renormalization of the disorder energies. This information may be
relevant to deal with interacting systems, since for many properties 
FSP models are a good approximation to the Anderson model and they 
can be extended to many-particle systems, whose direct simulation 
may be feasible for relatively large systems.

The distribution functions of the conductance in the FPS approximation 
and the Anderson model appear in many other
problems. Our results are
fully consistent with the strong version of the SPS \cite{SM01} if
we take into account that boundary conditions (leads) are not
irrelevant variables in the renormalization group sense, and may
change the universality class. We have checked that the skewness and 
the kurtosis obtained numerically for narrow leads tend to the 
theoretical predictions.  
We finally showed that the conductance distribution does not 
change when we apply a magnetic field.

The situation in 3D systems is more complex, since analytical 
approaches valid for 2D systems do not apply. There are no
hints of distribution functions for any similar type of problem.
Nevertheless, the quality of the fit in 2D systems suggests
that numerical simulations will provide useful information.

\section*{Acknowledgements}
 The authors would like to acknowledge financial
support from the Spanish DGI, project FIS2006-11126, and Fundación
Seneca, projects 03105/PI/05 and 05570/PD/07 (JP).

\end{document}